\documentclass[a4paper]{jpconf}
\usepackage{graphicx}
\usepackage[noadjust]{cite}
\begin{document}
\title{Transport in vertically stacked hetero-structures from 2D materials}
\author{Fan Chen, Hesameddin Ilatikhameneh, Yaohua Tan, Daniel Valencia, Gerhard Klimeck and Rajib Rahman}

\address{Network for Computational Nanotechnology (NCN), Purdue University, West Lafayette, IN 47906, USA}

\ead{fanchen@purdue.edu}

\begin{abstract}
In this work, the transport of tunnel field-effect transistor (TFET) based on vertically stacked hereto-structures from 2D transition metal dichalcogenide (TMD) materials is investigated by atomistic quantum transport simulations. WTe2-MoS2 combination was chosen due to the formation of a broken gap hetero-junction which is desirable for TFETs. There are two assumptions behind the MoS2-WTe2 hetero-junction tight binding (TB) model: 1) lattice registry. 2) The $S-Te$ parameters being the average of the $S-S$ and $Te-Te$ parameters of bilayer MoS2 and WTe2. The computed TB bandstructure of the hetero-junction agrees well with the bandstructure obtained from density functional theory (DFT) in the energy range of interest for transport. NEGF (Non-Equilibrium Green$'$s Function) equations within the tight binding description is then utilized for device transfer characteristic calculation.  Results show 1) energy filtering is the switching mechanism; 2) the length of the extension region is critical for device to turn off; 3) MoS2-WTe2 interlayer TFET can achieve a large on-current of $1000 \mu A/\mu m$ with $V_{DD} = 0.3V$, which suggests interlayer TFET can solve the low ON current problem of TFETs and can be a promising candidate for low power applications. 

\end{abstract}

\section{Introduction}

The fast growth of today$'$s technology has been sustained by continuous scaling of silicon-based MOSFETs. The scaling of transistors face two major challenges nowadays: the degradation of gate gate control and the fundamental thermionic limitation of the steepness of sub-threshold swing (SS). 2D materials have emerged as promising channel materials for transistors, as they can maintain excellent device electrostatics at much reduced channel length and thickness \cite{v23}. Interlayer tunnel field effect transistors (TFETs) based on vertically stacking 2D materials have also been shown to break the thermionic limitation of sub-threshold swing (SS)\cite{v11, v15}. However,  experimentally demonstrated vertical TFET with small SS still suffer from low current levels\cite{sarkar2015subthermionic}. TFET designs exploiting different 2D material properties\cite{ameen2015few, ilatikhameneh2016saving, chen2016thickness, 7301966, 1, chen2016configurable, chen2015transport} have been proposed to increase the tunneling current. Theoretical works also demonstrated the possibility of a high ON current in interlayer TFET$'$s \cite{v16, v17}. Yet, an understanding of device physics of interlayer TFETs such as the effect from device geometry, lattice mismatch, twisted angles\cite{tan2016first} and Van der waals bonding in order to provide a guideline for interlayer TFETs to achieve high ON currents experimentally. In this work, an interlayer TFET based on vertically stacked MoS2 and WTe2 as shown in Fig.~\ref{4}) is studied. The model assumptions are described in details and have been carefully examined. The switching mechanism and performance of the device are then investigated and evaluated. 
 \begin{figure}[h]
\begin{minipage}{11pc}
\includegraphics[width=11pc]{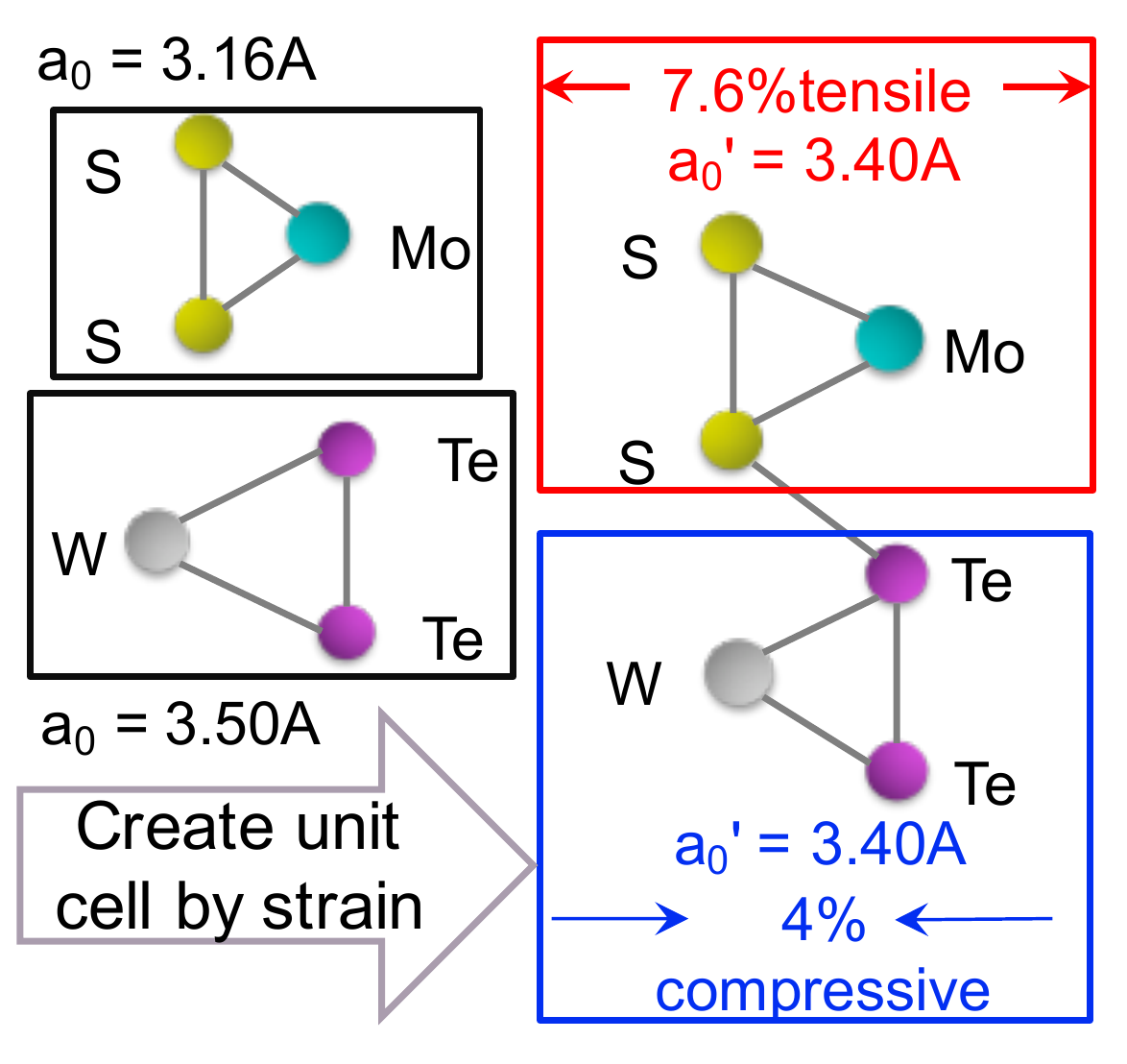}
\caption{\label{label}Assumption I : MoS2 and WTe2 layers have been strained to the same lattice constant to be registered.}
\label{1}
\end{minipage}\hspace{1pc}%
\begin{minipage}{11pc}
\includegraphics[width=10pc]{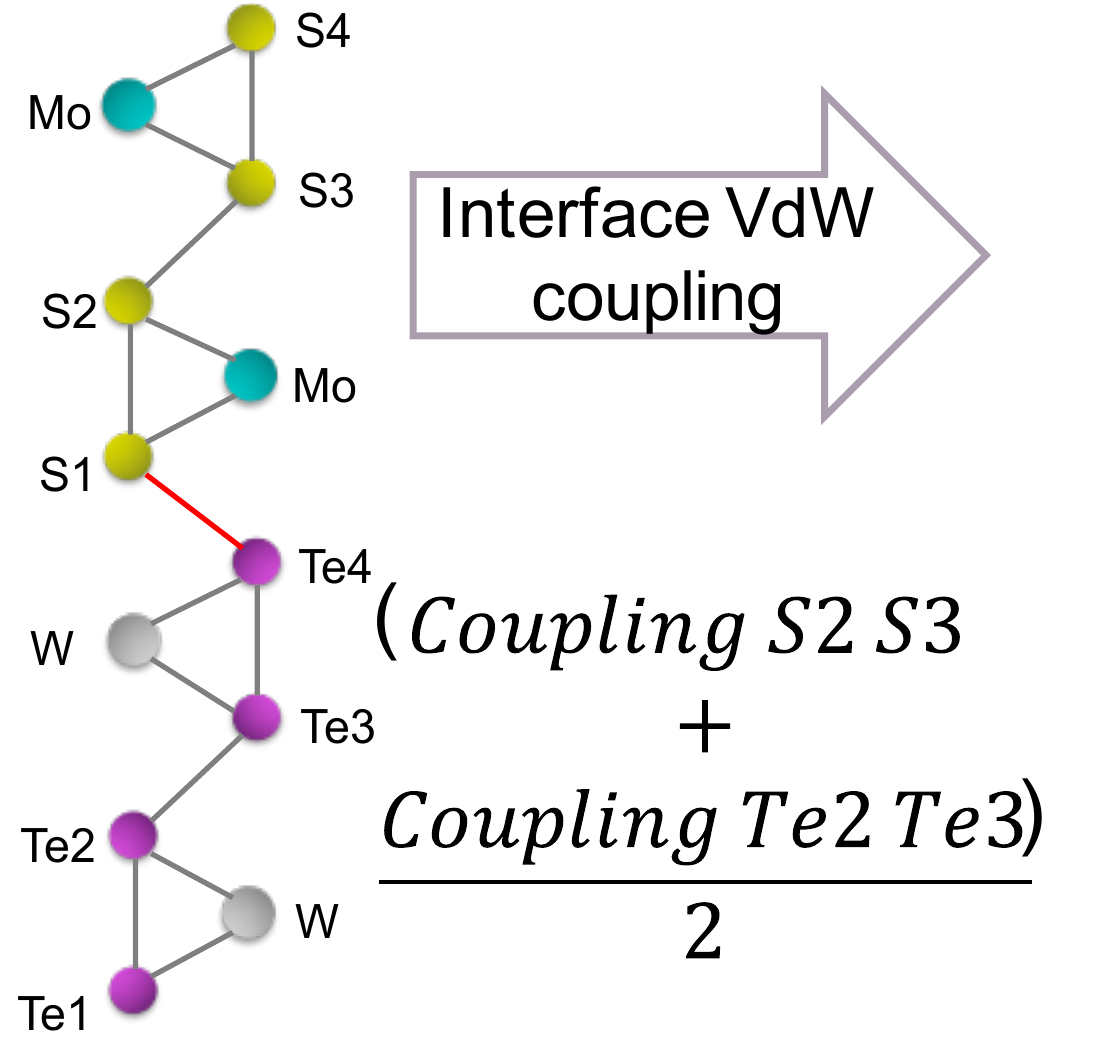}
\caption{\label{label}Assumption II : Interface VdW coupling of vertical MoS2-WTe2 hetero-junction is the average of the VdW coupling of the two materials.}
\label{2}
\end{minipage} \hspace{0.2pc}
\begin{minipage}{14pc}
\includegraphics[width=14pc]{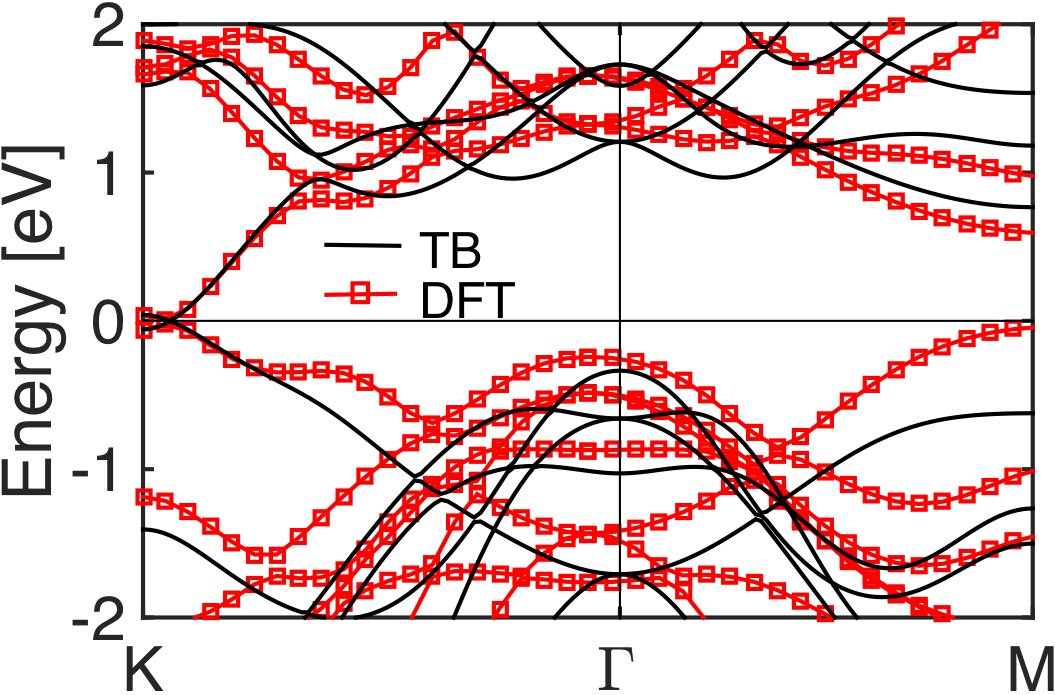}
\caption{\label{label}A comparison of band structure plots for vertical MoS2-WTe2 hetero-junction from tight binding (TB) and density function theory (DFT). They match well along the band edge at K point.}
\label{3}
\end{minipage} 
\end{figure}
\section{Method}
The Transistion Metal Dichalcogenide (TMD) Hamiltonian (strained) is represented by an $sp3d5$ second nearest neighbor tight-binding (TB) model including $spin-orbit$ coupling. DFT-guided TB model is adapted to reduce computational expense and system size compared to $ab-initio$ calculations \cite{tan2015tight}. The PBE functional is used in the density functional theory (DFT) calculations since it is known to produce band gap comparable with experimental measurements for TMDs\cite{21}. The Van der Waals correction for the interlayer coupling of multilayer MoS2 or WTe2 is included by  the optB88 functional\cite{farmanbar2016first}.The TB parameters are well calibrated to match the band structure and effective mass from DFT by a well-established mapping method \cite{tan2013empirical,tan2015tight}. The TB parameters are general and capture the band structure of both the bulk and monolayer TMDs \cite{21, fonseca2013efficient}. 

There are two major assumptions to obtain the TB parameters for the hetero-junction: 1) The materials are lattice matched by applying $7.6$ percent tensile strain to MoS2 and $4$ percent compressive strain to WTe2 as shown in Fig.~\ref{1}; 2) The interlayer coupling parameters between the Sulphur and the Telluride atoms are obtained by averaging the coupling parameters of S-S in bilayer MoS2 and and Te-Te in WTe2 as shown in Fig.~\ref{2}. 

A strained TB model is developed to fulfill the lattice registry. The relation of TB parameters with strain percentage is obtained by fitting to a set of strained DFT MoS2 or WTe2 monolayers. This model is then carefully examined by reproducing the DFT bandstructure under different strain. In order to obtain the DFT band structure of  MoS2-WTe2 interface for benchmarking, an ionic relaxation was carried out on the supercell until all the atomic forces on each ion were less than $0.0001 eV/nm$. Subsequently, the EK diagram was obtained for the relaxed configuration. This calculation was performed with a generalized gradient approximation (GGA) and PBE functional with a vdW correction as implemented in VASP (DFT-D2).

 All the transport characteristics of the MoS2-WTe2 TFET have been simulated using the self-consistent Poisson-semiclassical approach to obtain the potential profile and then this potential is passed to quantum transmitting boundary method (an equivalent of non-equilibrium Green$'$s function method for ballistic transport) in the multi-scale \cite{chen2015nemo5} and multi-physics \cite{miao2016buttiker, chen2015surface, 24} Nano-Electronic MOdeling (NEMO5) tool  \cite{fonseca2013efficient}. The default source and drain regions are doped with the doping level of $10^{20} cm^{-3}$. Equivalent oxide thickness (EOT) is set to 0.5nm.

\section{Results and Discussion}

 The DFT band-diagram of a relaxed MoS2-WTe2 vertical hetero-junction is plotted in Fig.~\ref{3}. The zero band gap in the band diagram shows MoS2 and WTe2 forms a broken bandgap junction. The conduction and valence band edges meet at K point which provides the density of states for carrier transport. Furthermore, the TB band diagram computed under the assumptions agrees well with the DFT results in the energy range that is important for transport. 
 
\begin{figure}[h]
\begin{minipage}{13pc}
\includegraphics[width=13pc]{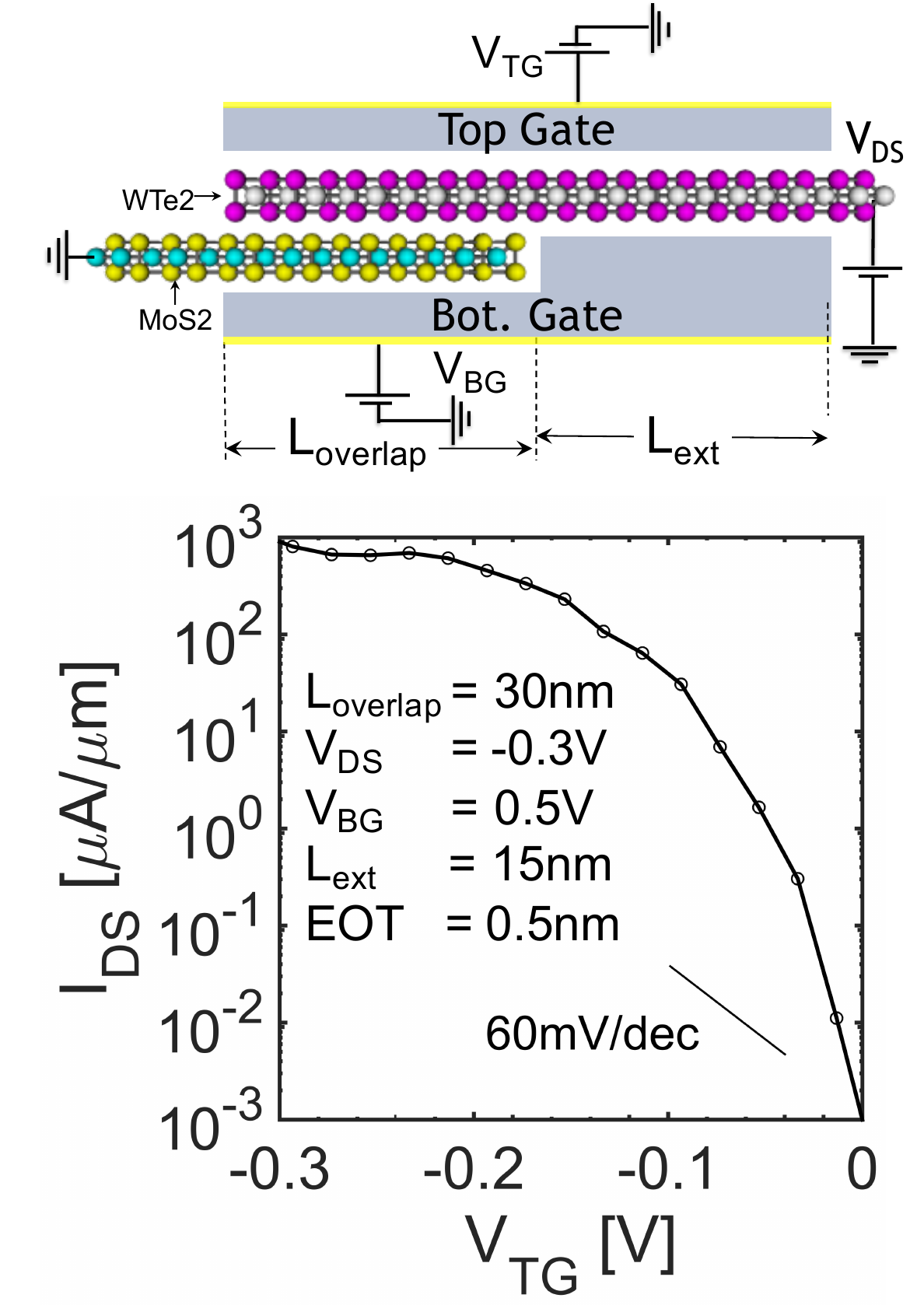}
\caption{\label{label}The device structure and the transfer characteristics ($Id-Vg$) of a MoS2-WTe2 interlayer TFET. The device has an overlap region and an extension region denoted by $L_{overlap}$ and $L_{ext}$, respectively.  }
\label{4}
\end{minipage}\hspace{3pc}%
\begin{minipage}{18pc}
\includegraphics[width=18pc]{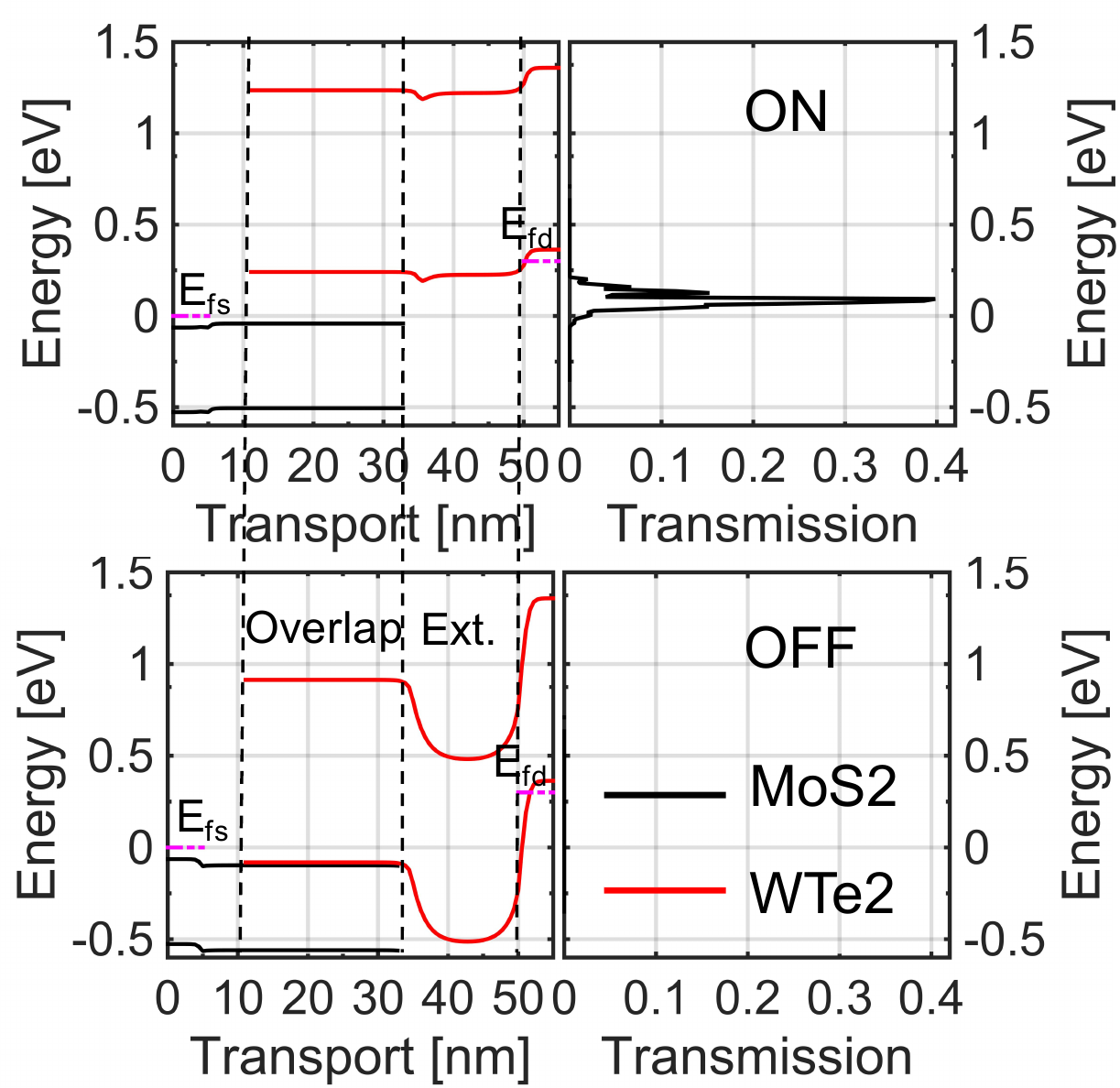}
\caption{\label{label} The band edges of a MoS2-WTe2 interlayer TFET at ON and OFF state aligned with the energy resolved transmission respectively. At ON state, WTe2 Ev is higher than MoS2 Ec. This opens a window for electrons to flow and it has a non-zero transmission. At OFF state, WTe2 Ev is lower than MoS2 Ec. No electrons can go through and it has a zero transmission.}
\label{5}
\end{minipage} 
\end{figure}

 The tight-binding material model developed in this manner was used to simulate quantum transport for the device shown in Fig.~\ref{4}.  The device geometry parameters of the MoS2-WTe2 interlayer TFET $L_{overlap}$ and $L_{ext}$ shown in Fig.~\ref{4} are optimized to be $30nm$ and $15nm$ respectively. The extension region is critical for the device to turn off. A zero $L_{ext}$ would cause a high leakage current such that the small SS can not be observed. The leakage current mainly comes from the edges states. Fig.~\ref{4} shows that the transfer characteristics for a supply voltage of -0.3 V.  The striking feature is the large ON current of $~1000 \mu A/\mu m$, which is significantly higher than the homo-junction 2D TFETs simulated thus far. The SS is around $20mV/dec$ in 4 orders of magnitude of current change. 

 The band edges of a MoS2-WTe2 interlayer TFET aligned with the energy resolved transmission at ON and OFF state respectively are plotted in Fig.~\ref{5}. At ON state, WTe2 Ev is higher than MoS2 Ec. This opens a window for electrons to flow, and therefore achieves a high ON current. This indicates an energy filtering switching mechanism. 


\section{Conclusion}

In conclusion, MoS2-WTe2 interlayer TFET is studied. The model assumptions have been carefully examined. The band diagram aligned with transmission shows the switching mechanism is the energy filtering. The MoS2-WTe2 interlayer TFET shows a SS smaller than $60mV/dec$. The extension region is critical for the device to turn off. An optimized device can achieve an ON current of $1000\mu A/\mu m$ with $V_{DD} = 0.3$.This suggests MoS2-WTe2 interlayer TFET as a promising candidate for low power applications.

\ack
This work was supported in part by the Center for Low Energy Systems Technology, one of six centers of STARnet, and in part by the Semiconductor Research Corporation Program through Microelectronics Advanced Research Corporation and Defense Advanced Research Projects Agency.\\

\section*{References}
\bibliography{all}

\end{document}